\documentclass{raa}          
\usepackage{graphicx,times}
\usepackage{natbib}
\usepackage{amssymb,amsmath}
\bibpunct{(}{)}{;}{a}{}{,}
\usepackage{longtable}

\usepackage[a4paper=true,dvipdfm=true,pagebackref=true]{hyperref}
\hypersetup{pdftitle = The title of my PDF, pdfauthor = My name, pdfsubject= The subject, pdfkeywords = keyword1 keyword2 keyword3} 
\hypersetup{colorlinks = true, linkcolor = green, anchorcolor = red, citecolor = blue, filecolor = red, pagecolor = red, urlcolor = red}

\begin{document}

   \title{Long term optical variability of bright X-ray point sources in elliptical galaxies}

 \volnopage{ {\bf 2013} Vol.\ {\bf X} No. {\bf XX}, 000--000}
   \setcounter{page}{1}

   \author{V. Jithesh\inst{1}, R. Misra\inst{2},
      P. Shalima\inst{2}, K. Jeena\inst{3}, C. D. Ravikumar\inst{1} and B. R. S. Babu\inst{1,4}
   }

   \institute{ Department of Physics, University of Calicut, Malappuram-673635, India; {\it jitheshthejus@gmail.com}\\
        \and
             Inter-University Centre for Astronomy and Astrophysics, Post Bag4, Ganeshkhind, Pune-411007, India; {\it rmisra@iucaa.ernet.in}\\
	\and
Department of Physics, Providence Women's College, Malaparamba, Calicut-673009, India\\
\and 
Department of Physics, Sultan Qaboos University, Muscat, Oman\\
}

\abstract{ We present long term optical variability studies of bright X-ray
  sources in four nearby elliptical galaxies with {\it Chandra}
  Advanced CCD Imaging Spectrometer array (ACIS-S) and {\it Hubble
  Space Telescope (HST)} Advanced Camera for Surveys observations. Out
  of the 46 bright (X-ray counts $> 60$) sources that are in the
  common field of view of the {\it Chandra} and {\it HST} observations, 34
  of them have potential optical counterparts, while the rest of them
  are optically dark. After taking into account of systematic errors,
  estimated using the field optical sources as reference, we
  find that four of the X-ray sources (three in NGC1399 and one in
  NGC1427) have variable optical counterparts at a high significance
  level. The X-ray luminosities of these source are $\sim 10^{38}$
  $\rm ergs~s^{-1}$ and are also variable on similar time-scales. The optical
  variability implies that the optical emission is associated with
  the X-ray source itself rather than being the integrated light from
  a host globular cluster. For one source the change in optical magnitude is $> 0.3$, which is one of the
  highest reported for this class of X-ray sources and this suggests
  that the optical variability is induced by the X-ray activity. However,
  the optically variable sources in NGC1399 have been reported
  to have blue colours ($g - z > 1$). All four sources have been detected in the
  infra-red (IR) by {\it Spitzer} as point sources, and their 
ratio of $5.8$ to $3.6 \mu m$ flux are $> 0.63$ indicating that their IR
spectra are like those of Active Galactic Nuclei (AGN). 
While spectroscopic confirmation is required, it is likely that
all four sources are background AGNs. We find none of the X-ray sources 
having optical/IR colours different from AGNs, to be optically variable.
}

   \authorrunning{V. Jithesh et al. }            
   \titlerunning{Long term optical variability of bright X-ray point sources in elliptical galaxies}  
   \maketitle

%
\section{Introduction}           
\label{sect:intro}

The unprecedented angular resolution of {\it Chandra} satellite has enabled the study of
X-ray point sources in nearby galaxies. Most of these point sources are expected to
be X-ray binaries like the ones found in the Milky Way. An important result of the
{\it Chandra} observations was the confirmation of 
Ultra-luminous X-ray sources (ULXs), discovered with { \it Einstein} observatory 
in the 1980s \citep{Fab89}. These are off-nuclear X-ray point sources 
with X-ray luminosities in the range $10^{39}-10^{41}$ $\rm ergs~s^{-1}$. The observed luminosities 
of ULXs exceed the Eddington limit for a $10 M_{\odot}$ black hole, which has 
led to a sustained debate on the nature of these sources. Since ULXs are off-nuclear 
sources, their masses must be $< 10^{5}M_{\odot}$ from dynamical friction arguments 
\citep{Kaa01}. Thus, ULXs may represent a class of Intermediate Mass Black holes (IMBHs) 
whose mass range ($10 M_{\odot} < M < 10^{5}M_{\odot}$) is between that of stellar 
mass black holes and super massive ones \citep{Mak00}. 
Further the nature of the sources in nearby galaxies, which are
less luminous than ULX, is also not clear and it is 
difficult to ascertain whether they harbour neutron stars or black holes.

The primary reason for these uncertainties is that unlike Galactic X-ray binaries, it is difficult
to identify the companion star in the optical and obtain the binary parameters. For most X-ray sources
in nearby galaxies, the associated optical emission is due to the integrated light from a host globular
cluster \citep{Kim06,Kim09,Pta06,Goa02} and it is usually not possible to resolve and identify the companion
star. However, these studies provide important information regarding the environment of the X-ray sources. For example,
ULXs in early type galaxies are associated with red globular clusters \citep{Pta06, Ang01}. Even the non-detection
of optical emission allows one to impose 
strong upper limit on the black hole mass for these accreting systems based on some 
standard assumptions \citep{Jit11}. However, a more direct inference on the nature of the system
requires identification and spectral measurement of the associated optical emission. An important aspect of
identifying the correct optical counterpart in a crowded field is to check for optical variability.
If the optical emission is variable, it is most probably directly associated with the X-ray source
and not the integrated light of stars in a globular cluster. Indeed, for low mass X-ray binaries
in the Galaxy, the optical emission is variable and is for some cases correlated with the X-ray
emission \citep[e.g.4U 1636-536:][]{Shi11} while for others it is not \citep[e.g. GX 9+9:][]{Kon06}.
The optical variability may be due to the orbital motion of the donor star or reprocessing of the
variable X-ray emission or X-ray heating of the companion. However, typically the optical counterpart 
of X-ray binaries in nearby galaxies will not be resolved, especially if the source is in a globular cluster. Hence
it is not expected that optical variability will be seen for them.

Nevertheless, variability of optical counterparts have been measured for the bright X-ray sources
in nearby galaxies. For example, 
the optical counterpart of NGC1313 X-2 has been identified as a O7 star at solar metallicity, 
The optical counterpart exhibits variability at $\sim$ 0.2 mag on short time scales \citep{Liu07, Gri08} and the
variability may be due to varying X-ray irradiation of the donor star and a stochastic varying 
contribution from the accretion disk. 
An independent study of the same source \citep{Muc07} revealed that the optical flux of the counterpart shows
variation ${\leq} 30$\% and that it may be a main-sequence star of mass $\sim$ $10-18 M_\odot$ feeding to a 
black hole of mass $120 M_\odot$. The optical counterpart of Holmberg IX X-1 exhibits photometric variability of $0.136\pm0.027$ in the {\it HST/ACS V} band images \citep{Gri11} although it seems to have a constant magnitude within 
photometric errors ($22.710\pm0.038$ and $22.680\pm0.015$) in SUBARU {\it V} band images. 
\cite{Tao11} have reported the optical variability for three ULXs, M101 ULX-1, M81 ULX1 and NGC1313 X-2, at 
a magnitude difference of 0.2 or larger in the {\it V} band. 
Some of the X-ray sources in nearby galaxies could
be background AGN and it is expected that their optical emission would be variable.

It is important to identify more X-ray sources that have optically variable counterparts, which then can be subjected to
more detailed observational follow-ups such as spectral and/or simultaneous X-ray/optical
observations. A systematic analysis of a number of galaxies to identify such sources will be crucial to understand the nature of these sources.
Such an analysis would require multiple optical observations of a galaxy, a uniform scheme to
identify optical counterparts of the X-ray sources and more importantly an estimate of
the systematic uncertainties in order to avoid any spurious variability that may arise
if only statistical errors are considered.

In this work, we consider elliptical galaxies which are $\lesssim 20$ Mpc away that
have been observed by {\it Chandra} and have more than one {\it HST} observation in the same
filter. We restrict our analysis to ellipticals since for them the continuum
optical emission can be modelled and subtracted out to reveal optical point sources
\citep{Jit11}. Using the field optical sources we estimate the systematic errors in the
optical flux measurements and hence can report true optical variability at a high
confidence level. Our aim is to study the optical counterpart variability of bright X-ray sources
(X-ray counts $> 60$) whose X-ray spectra can be modelled and hence a reliable estimate
of its luminosity can be obtained.
 
In the next section, we describe the selection of the sample galaxies. \S 3 and \S 4 describe the 
X-ray analysis and the method to identify the optical counterparts and to compute the 
photometry with systematic errors. We discuss the results in \S 5. 

\section{Source Selection}
\label{sect:select}

The samples were selected based on three criteria. (1) The distance to the host 
galaxy is $\lesssim 20$ Mpc, (2) The galaxy has {\it Chandra} observation and (3) 
has more than one epoch {\it HST} observation in the same filter. Based on these criteria, we have selected five 
galaxies which are listed in Table \ref{sample}. For three of the galaxies there are multiple
{\it Chandra} observations which we use to study the long term X-ray variability.
Using the longest exposure {\it Chandra} observations we identify X-ray sources which
have counts $> 60$, so that we can obtain reliable X-ray spectra for them. Of these, we 
selected those that fell within the field of view of both the {\it HST} observations. For NGC2768
the only source that fulfilled these criteria was the central AGN and hence we report no further
analysis of the galaxy.

\begin{table}
\bc
\begin{minipage}[]{100mm}
\caption[]{Sample Galaxy Properties\label{sample}}\end{minipage}
\setlength{\tabcolsep}{2.0pt}
\small
 \begin{tabular}{cccccccccc}
  \hline\noalign{\smallskip}
Galaxy & Distance & {\it Chandra} & {\it Chandra} & $T_{exp}$ & {\it HST} & {\it HST} & {\it HST} & $N_{xo}$\\
&(Mpc)&ID&Observation Date&(ks)&ID&Filter&Observation Date&&\\
  \hline\noalign{\smallskip}
NGC1399 & $18.9$ & $9530$ & $2008Jun08$ & $60.11$ & $J9P305020$ & $F475W$ & $2006Aug02$ & 18 \\
	&  	 & $319$  & $2000Jan18$ & $56.66$ & $J90X02020$ & $F475W$ & $2004Sep11$ &  \\
NGC4486 & $15.8$ & $2707$ & $2002Jul06$ & $99.93$ & $J9E086010$ & $F814W$ & $2006Feb20$ & 17 \\
	&  	 & $352$  & $2000Jul29$ & $38.16$ & $J9E003010$ & $F814W$ & $2006Jan03$ &    \\
NGC4278 & $15.2$ & $7081$ & $2007Feb20$ & $112.14$ & $J9NM06010$ & $F850LP$ & $2007Jan02$ & 9 \\
	&  	 & $11269$ & $2010Mar15$ & $82.95$ & $J9NM07010$ & $F850LP$ & $2006Dec23$ &   \\ 
NGC1427 & $21.1$ & $4742$ & $2005May01$ & $51.70$ & $J9P302020$ & $F475W$ & $2006Jul31$ & 2 \\
	&  	 & $-$    & $-$     & $-$ & $J90X06020$ & $F475W$ & $2004Sep22$ &  \\
NGC2768 & $20.1$ & $9528$ & $2008Jan25$ & $65.46$ & $J6JT08021$ & $F814W$ & $2002May31$ & 1 \\
	&  	 & $-$    & $-$     & $-$ & $J8DT02021$ & $F814W$ & $2003Jan14$ &  \\
  \noalign{\smallskip}\hline
\end{tabular}
\ec
\tablecomments{0.92\textwidth}{(1) Host galaxy name; (2) Distance to the host galaxy from NED; (3) {\it Chandra} observation ID; 
(4) Chandra Observation Date; (5) Exposure time in kilo seconds; (6) {\it HST} observation ID; (7) {\it HST} Filter; (8) Observation Date; 
(9) Number of common sources in the field of view of X-ray and optical images.}
\end{table}

NGC1399 and NGC4486 are giant elliptical galaxies in the center of the 
Fornax and Virgo clusters respectively and are well-known for their populous 
globular cluster systems \citep{Kim06,Dir03,Bas06,Ang01,Jor04,Irw06,Siv07}. 
The {\it Chandra} analysis \citep{Ang01} of NGC1399 shows that a large fraction 
of $2-10{\rm~keV}$ X-ray emission is most likely from the low-mass X-ray binaries 
(LMXBs). The {\it HST} study of these {\it Chandra} 
identified X-ray sources shows that $\sim70$\% (26 of 38 sources) of these sources 
are associated to Globular Clusters (GCs). The specific frequency of globular cluster 
in this galaxy is 2-3 times that of typical elliptical galaxies \citep{Har91}. 
The optical counterparts of the ULXs (CXOJ033831.8-352604) show [OIII] ${\lambda}5007$ and [NII] ${\lambda}6583$ 
emission line in the optical spectrum \citep{Irw10}. \cite{Irw10} suggest that the lack of H${\alpha}$ and 
H${\beta}$ emission line in the spectrum may be an indication of a disruption of a white dwarf star by an intermediate 
mass black hole (IMBH).  
 
The analysis of {\it Chandra} deep observations of the nearby elliptical galaxy NGC4278, 
identified 236 X-ray point sources with luminosity ranging from $3.5\times10^{36} \rm ergs~s^{-1}$ to $2\times10^{40} \rm ergs~s^{-1}$ \citep{Bra09}. This galaxy has rich 
globular cluster systems and 39 of them are coincident with X-ray sources which lie within the $D_{25}$ 
ellipse of the galaxy. 10 of the GC-LMXB associated sources lie at the high X-ray luminosity end ($L_{X} 
> 10^{38} \rm ergs~s^{-1}$). Also, $44$\% of the X-ray source 
population exhibit long term variability indicating that they are accreting 
compact objects. \cite{Fab10} analysed the spectra of the X-ray sources by 
fitting with either single thermal accretion disk or power law model and the best-fit 
parameters are similar to those of Galactic BH binaries. Seven luminous sources have luminosity 
exceeding the Eddington limit for accreting neutron stars. Four of these sources are 
associated with GCs and the other three do not have optical counterparts and are found 
in the stellar field of NGC4278.

NGC1427 is a low luminosity elliptical galaxy in Fornax cluster and its globular cluster 
association has been studied by \cite{For01} and \cite{Kis97}. The photometry studies reveal a bimodal 
cluster population in this galaxy and suggest that the formation mechanism of globular clusters 
in low luminosity galaxies shows similarities with giant galaxies. The {\it Chandra} ACIS 
Survey of X-ray point sources \citep{Liu11} identified two ULXs in this galaxy with luminosity 
$\ge 2\times10^{39} \rm ergs~s^{-1}$. Among them, one source is inside the $D_{25}$ region of the galaxy, 
and the other is outside the $D_{25}$ region.
 
\section{X-ray Analysis}
\label{sect:xray}

We start with analysing the {\it Chandra} observations listed in Table \ref{sample}.
These are observations with Advanced CCD Imaging Spectrometer array 
(ACIS-S) and the data reduction and analysis were done using {\sc ciao 4.2}, and 
{\sc heasoft 6.9}. Using the {\sc ciao} source detection tool {\it celldetect}, the X-ray point 
sources were extracted from the level 2 event list with {\it signal-to-noise} ratio of 3. 
Some of the extracted sources are near the nucleus and in the excessive diffused emission 
regions and hence these sources were not included in the analysis. The extracted sources with net count ${\geq} 60$ were selected. 
The spectral analysis was done using {\sc xspec 12.6.0}, and the data were fitted in the energy range of 0.3 - 8.0 keV. 

All sources were fitted with two spectral models: an absorbed power law and an absorbed disk black body. 
Absorption was taken into account using the {\sc xspec} model {\it wabs}. If the $\chi^2$ difference between
the two models was larger than 2.7, we took the model with the smaller $\chi^2$ to be the representative
one. If the $\chi^2$ difference was less than 2.7 (i.e. when both models equally well represent the data),
we choose the representative model to be the one which gave a lower luminosity. The analysis has been done for both observations
listed in Table \ref{sample}, with the longer observation being called the first one and the shorter one the second.
Table \ref{firstsecondfavour} lists the spectral parameters corresponding to the representative model. 
The spectra of two sources in NGC1399 are not well fitted with either model and a closer
inspection revealed the presence of an additional {\it mekal} component which has been added.

To quantify the long term variability of the X-ray sources we consider sources that are in the
field of view of both observations. We jointly fit the spectra using the same model parameters except
that we introduce a constant factor which multiplies the later observation. In other words, we keep 
the absorption and the spectral parameters (i.e. either the temperature or the power-law index) same
for both data sets, but allow for variation in the relative normalization. If the constant is unity, then
the source has not varied. We consider a source to be X-ray variable only if the constant 
$C_{2}$ is inconsistent with unity at 2-sigma level i.e. $|C_{2}-1|/\sigma_{C_{2}} > 2$. The results of the joint
fitting are shown in Table \ref{xrayvariable}. As expected, several of the X-ray sources clearly
exhibit long term variability.

\begin{table}
\bc
\begin{minipage}[]{150mm}
\caption[]{Spectral Properties of point sources and best-fit models for first and second epoch\label{firstsecondfavour}}\end{minipage}
\setlength{\tabcolsep}{3.0pt}
\scriptsize
 \begin{tabular}{ccccccccccccc}
  \hline\noalign{\smallskip}
Galaxy & RA (J2000) & Dec (J2000) & $n_H$ & $\Gamma/kT_{in}$ & log($L_{1}$) & $\chi^2 / d.o.f$ & Model & $n_H$ & $\Gamma/kT_{in}$ & log($L_{2}$) & $\chi^2 / d.o.f$ & Model\\
  \hline\noalign{\smallskip}

NGC1399 & 3h38m32.58s & -35\dg27$'$5.40$"$ & $0.03^{+ 0.10}_{- 0.03}$ & $1.63^{+ 0.33}_{- 0.21}$ & $39.39^{+ 0.06}_{- 0.05}$ & $11.25/22$ & P & $0.03^{+ 0.04}_{- 0.03}$ & $1.61^{+ 0.20}_{- 0.19}$ & $39.53^{+ 0.04}_{- 0.04}$ & $47.68/39$ & P \\
NGC1399 & 3h38m31.79s & -35\dg26$'$4.23$"$ & $0.01^{+ 0.07}_{- 0.01}$ & $0.33^{+ 0.05}_{- 0.06}$ & $39.04^{+ 0.15}_{- 0.06}$ & $ 8.30/13$ & D & $0.00^{+ 0.01}_{- 0.00}$ & $0.38^{+ 0.04}_{- 0.03}$ & $39.10^{+ 0.04}_{- 0.04}$ & $23.68/27$ & D \\  
NGC1399 & 3h38m36.82s & -35\dg27$'$46.98$"$ & $0.00^{+ 0.14}_{- 0.00}$ & $1.67^{+ 0.61}_{- 0.30}$ & $38.72^{+ 0.10}_{- 0.11}$ & $ 4.39/4$ & P & $0.00^{+ 0.08}_{- 0.00}$ & $2.27^{+ 0.74}_{- 0.30}$ & $38.78^{+ 0.14}_{- 0.09}$ & $ 9.73/9$ & P \\
NGC1399 & 3h38m33.09s & -35\dg27$'$31.53$"$ & $0.00^{+ 0.13}_{- 0.00}$ & $0.61^{+ 0.18}_{- 0.17}$ & $38.61^{+ 0.12}_{- 0.08}$ & $ 5.91/6$ & D & $0.00^{+ 0.04}_{- 0.00}$ & $1.73^{+ 0.31}_{- 0.23}$ & $38.97^{+ 0.09}_{- 0.10}$ & $20.47/11$ & P \\
NGC1399 & 3h38m25.95s & -35\dg27$'$42.19$"$ & $0.00^{+ 0.16}_{- 0.00}$ & $1.05^{+ 0.98}_{- 0.40}$ & $38.64^{+ 0.18}_{- 0.15}$ & $ 3.41/4$ & D & $0.00^{+ 0.14}_{- 0.00}$ & $0.87^{+ 0.63}_{- 0.29}$ & $38.55^{+ 0.15}_{- 0.14}$ & $ 3.48/3$ & D \\
NGC1399 & 3h38m32.76s & -35\dg26$'$58.73$"$ & $0.00^{+ 0.18}_{- 0.00}$ & $2.58^{+ 0.00}_{- 1.24}$ & $38.86^{+ 0.19}_{- 0.19}$ & $ 6.13/5$ & D & $0.00^{+ 0.08}_{- 0.00}$ & $1.63^{+ 0.59}_{- 0.39}$ & $38.74^{+ 0.16}_{- 0.18}$ & $ 5.39/7$ & P \\
NGC1399 & 3h38m32.34s & -35\dg27$'$2.11$"$ & $0.99^{+ 1.41}_{- 0.67}$ & $0.68^{+ 0.50}_{- 0.27}$ & $38.94^{+ 0.52}_{- 0.25}$ & $ 2.53/5$ & D & $0.62^{+ 0.00}_{- 0.00}$ & $1.33^{+ 0.00}_{- 1.06}$ & $38.39^{+ 3.21}_{- 0.57}$ & $ 3.00/5$ & D \\
NGC1399 & 3h38m31.86s & -35\dg26$'$49.26$"$ & $0.10^{+ 1.76}_{- 0.10}$ & $0.84^{+ 0.00}_{- 0.68}$ & $38.41^{+ 0.58}_{- 0.31}$ & $ 3.72/4$ & D & $0.00^{+ 0.09}_{- 0.00}$ & $2.48^{+ 0.84}_{- 0.37}$ & $38.76^{+ 0.20}_{- 0.10}$ & $11.90/9$ & P \\
\#NGC1399 & 3h38m25.66s & -35\dg27$'$41.50$"$ & $0.00^{+ 0.17}_{- 0.00}$ & $1.09^{+ 7.98}_{- 0.50}$ & $38.67^{+ 0.37}_{- 0.18}$ & $ 6.71/4$ & D & $-$ & $-$ & $<38.08$ & $-$ & $-$ \\
\#NGC1399 & 3h38m27.80s & -35\dg25$'$26.65$"$ & $0.00^{+ 0.71}_{- 0.00}$ & $1.27^{+ 1.30}_{- 0.69}$ & $38.55^{+ 0.17}_{- 0.18}$ & $ 0.53/2$ & D & $-$ & $-$ & $<38.25$ & $-$ & $-$ \\
*NGC1399 & 3h38m26.50s & -35\dg27$'$32.29$"$ & $-$ & $-$ & $<38.08$ & $-$ & $-$ & $0.00^{+ 0.09}_{- 0.00}$ & $0.90^{+ 0.34}_{- 0.25}$ & $38.71^{+ 0.11}_{- 0.11}$ & $ 8.29/7$ & D \\
*NGC1399 & 3h38m33.82s & -35\dg25$'$56.95$"$ & $-$ & $-$ & $<37.94$ & $-$ & $-$ & $0.00^{+ 0.10}_{- 0.00}$ & $0.48^{+ 0.43}_{- 0.18}$ & $38.35^{+ 0.14}_{- 0.14}$ & $ 5.25/3$ & D \\
*NGC1399 & 3h38m33.80s & -35\dg26$'$58.30$"$ & $-$ & $-$ & $<38.73$ & $-$ & $-$ & $0.54^{+ 1.26}_{- 0.54}$ & $0.20^{+ 0.62}_{- 0.12}$ & $39.03^{+ 2.95}_{- 1.06}$ & $ 0.44/2$ & D \\
*NGC1399 & 3h38m32.35s & -35\dg27$'$10.63$"$ & $-$ & $-$ & $<38.16$ & $-$ & $-$ & $0.03^{+ 0.24}_{- 0.03}$ & $0.98^{+ 0.62}_{- 0.36}$ & $38.63^{+ 0.12}_{- 0.14}$ & $ 0.08/6$ & D \\
*NGC1399 & 3h38m25.32s & -35\dg27$'$53.49$"$ & $-$ & $-$ & $<38.20$ & $-$ & $-$ & $0.00^{+ 0.22}_{- 0.00}$ & $1.86^{+ 3.53}_{- 0.82}$ & $38.62^{+ 0.15}_{- 0.17}$ & $ 2.20/3$ & D \\
*NGC1399 & 3h38m27.19s & -35\dg26$'$1.53$"$ & $-$ & $-$ & $<38.38$ & $-$ & $-$ & $0.00^{+ 0.33}_{- 0.00}$ & $1.28^{+ 1.59}_{- 0.68}$ & $38.71^{+ 0.25}_{- 0.23}$ & $ 5.29/3$ & P \\
\ddag NGC1399 & 3h38m27.63s & -35\dg26$'$48.54$"$ & $0.15^{+ 0.12}_{- 0.11}$ & $2.72^{+ 0.63}_{- 0.52}$ & $39.48^{+ 0.26}_{- 0.15}$ & $23.83/22$ & P & $0.01^{+ 0.03}_{- 0.01}$ & $0.42^{+ 0.05}_{- 0.06}$ & $39.15^{+ 0.07}_{- 0.05}$ & $30.26/30$ & D \\
\ddag NGC1399 & 3h38m38.76s & -35\dg25$'$54.86$"$ & $0.00^{+ 0.10}_{- 0.00}$ & $1.06^{+ 0.30}_{- 0.26}$ & $38.97^{+ 0.08}_{- 0.08}$ & $12.31/10$ & D & $0.00^{+ 0.29}_{- 0.00}$ & $2.02^{+ 1.97}_{- 0.56}$ & $38.56^{+ 0.58}_{- 0.14}$ & $ 3.71/2$ & P \\

NGC4486 & 12h30m47.15s & 12\dg24$'$15.91$"$ & $0.00^{+ 0.01}_{- 0.00}$ & $0.66^{+ 0.08}_{- 0.07}$ & $39.17^{+ 0.04}_{- 0.04}$ & $108.78/83$ & D & $0.25^{+ 0.14}_{- 0.12}$ & $2.91^{+ 0.73}_{- 0.57}$ & $39.75^{+ 0.33}_{- 0.19}$ & $65.87/44$ & P \\
NGC4486 & 12h30m53.24s & 12\dg23$'$56.69$"$ & $0.03^{+ 0.11}_{- 0.03}$ & $1.05^{+ 0.26}_{- 0.21}$ & $39.03^{+ 0.07}_{- 0.08}$ & $85.86/72$ & D & $0.00^{+ 0.09}_{- 0.00}$ & $0.95^{+ 0.36}_{- 0.27}$ & $38.96^{+ 0.10}_{- 0.12}$ & $50.16/34$ & D \\
NGC4486 & 12h30m50.12s & 12\dg23$'$1.07$"$ & $0.00^{+ 0.07}_{- 0.00}$ & $1.11^{+ 0.33}_{- 0.24}$ & $38.97^{+ 0.08}_{- 0.10}$ & $88.36/84$ & D & $0.00^{+ 0.50}_{- 0.00}$ & $0.60^{+ 0.50}_{- 0.39}$ & $38.69^{+ 0.75}_{- 0.25}$ & $32.89/40$ & D \\
NGC4486 & 12h30m46.19s & 12\dg23$'$28.63$"$ & $0.00^{+ 0.07}_{- 0.00}$ & $0.92^{+ 0.23}_{- 0.19}$ & $38.95^{+ 0.07}_{- 0.08}$ & $76.43/70$ & D & $0.01^{+ 0.17}_{- 0.01}$ & $0.96^{+ 0.46}_{- 0.38}$ & $38.99^{+ 0.11}_{- 0.13}$ & $36.06/30$ & D \\
NGC4486 & 12h30m44.67s & 12\dg22$'$1.06$"$ & $0.25^{+ 0.23}_{- 0.15}$ & $2.62^{+ 1.07}_{- 0.69}$ & $39.16^{+ 0.47}_{- 0.21}$ & $51.11/48$ & P & $0.06^{+ 0.39}_{- 0.06}$ & $1.21^{+ 0.00}_{- 0.60}$ & $38.82^{+ 0.31}_{- 0.23}$ & $21.96/22$ & D \\
NGC4486 & 12h30m50.80s & 12\dg25$'$2.00$"$ & $0.00^{+ 0.09}_{- 0.00}$ & $1.21^{+ 0.54}_{- 0.34}$ & $38.82^{+ 0.10}_{- 0.12}$ & $59.29/46$ & D & $0.02^{+ 0.25}_{- 0.02}$ & $1.58^{+ 1.61}_{- 0.60}$ & $39.04^{+ 0.13}_{- 0.15}$ & $17.77/16$ & D \\
NGC4486 & 12h30m44.26s & 12\dg22$'$9.37$"$ & $0.00^{+ 0.29}_{- 0.00}$ & $0.49^{+ 0.44}_{- 0.28}$ & $38.36^{+ 0.46}_{- 0.19}$ & $65.61/40$ & D & $0.00^{+ 0.35}_{- 0.00}$ & $1.62^{+ 0.00}_{- 0.84}$ & $38.69^{+ 0.23}_{- 0.29}$ & $25.09/19$ & D \\
\#NGC4486 & 12h30m44.71s & 12\dg24$'$34.61$"$ & $0.00^{+ 0.04}_{- 0.00}$ & $2.11^{+ 0.33}_{- 0.15}$ & $39.08^{+ 0.06}_{- 0.05}$ & $58.46/55$ & P & $-$ & $-$ & $<38.52$ & $-$ & $-$ \\
\#NGC4486 & 12h30m46.32s & 12\dg23$'$23.19$"$ & $0.00^{+ 0.12}_{- 0.00}$ & $0.65^{+ 0.16}_{- 0.20}$ & $38.89^{+ 0.11}_{- 0.08}$ & $94.05/68$ & D & $-$ & $-$ & $<38.51$ & $-$ & $-$ \\
\#NGC4486 & 12h30m47.32s & 12\dg23$'$8.82$"$ & $0.02^{+ 0.19}_{- 0.02}$ & $0.76^{+ 0.26}_{- 0.33}$ & $38.84^{+ 0.15}_{- 0.11}$ & $95.28/80$ & D & $-$ & $-$ & $<38.58$ & $-$ & $-$ \\
\#NGC4486 & 12h30m50.08s & 12\dg22$'$51.21$"$ & $0.00^{+ 0.15}_{- 0.00}$ & $0.66^{+ 0.69}_{- 0.44}$ & $38.46^{+ 0.13}_{- 0.27}$ & $69.39/69$ & D & $-$ & $-$ & $<38.63$ & $-$ & $-$ \\
\#NGC4486 & 12h30m52.79s & 12\dg23$'$36.85$"$ & $3.38^{+ 1.38}_{- 0.63}$ & $9.50^{+ 0.00}_{- 12.50}$ & $44.10^{+ 5.62}_{- 2.00}$ & $73.37/69$ & P & $-$ & $-$ & $<44.78$ & $-$ & $-$ \\
\#NGC4486 & 12h30m43.49s & 12\dg23$'$46.80$"$ & $0.04^{+ 0.72}_{- 0.04}$ & $0.71^{+ 0.77}_{- 0.48}$ & $38.34^{+ 0.74}_{- 0.28}$ & $23.79/32$ & D & $-$ & $-$ & $<38.41$ & $-$ & $-$ \\
\#NGC4486 & 12h30m46.52s & 12\dg24$'$50.15$"$ & $0.00^{+ 0.38}_{- 0.00}$ & $0.70^{+ 0.65}_{- 0.47}$ & $38.40^{+ 0.50}_{- 0.19}$ & $36.82/32$ & D & $-$ & $-$ & $<38.41$ & $-$ & $-$ \\
\#NGC4486 & 12h30m44.91s & 12\dg24$'$4.50$"$ & $0.00^{+ 0.83}_{- 0.00}$ & $3.13^{+ 0.00}_{- 2.20}$ & $38.43^{+ 0.19}_{- 0.27}$ & $32.28/38$ & D & $-$ & $-$ & $<38.69$ & $-$ & $-$ \\
\#NGC4486 & 12h30m50.82s & 12\dg24$'$11.80$"$ & $0.08^{+ 0.16}_{- 0.08}$ & $0.50^{+ 0.19}_{- 0.16}$ & $38.80^{+ 0.19}_{- 0.15}$ & $55.48/67$ & D & $-$ & $-$ & $<38.55$ & $-$ & $-$ \\
\#NGC4486 & 12h30m49.13s & 12\dg21$'$59.40$"$ & $0.00^{+ 65.00}_{- 36.13}$ & $0.58^{+ 0.00}_{- 3.58}$ & $38.69^{+ 17.17}_{- 9.11}$ & $60.72/41$ & P & $-$ & $-$ & $<38.88$ & $-$ & $-$ \\

NGC4278 & 12h20m7.75s & 29\dg17$'$20.39$"$ & $0.00^{+ 0.07}_{- 0.00}$ & $1.71^{+ 0.64}_{- 0.42}$ & $38.64^{+ 0.08}_{- 0.09}$ & $7.36/11$ & D & $0.00^{+ 0.11}_{- 0.00}$ & $1.46^{+ 0.92}_{- 0.44}$ & $38.61^{+ 0.13}_{- 0.12}$ & $12.44/7$ & D \\
NGC4278 & 12h20m3.43s & 29\dg16$'$39.35$"$ & $0.00^{+ 0.14}_{- 0.00}$ & $1.71^{+ 1.24}_{- 0.55}$ & $38.49^{+ 0.13}_{- 0.13}$ & $4.40/6$ & D & $0.00$ & $1.22$ & $38.26$ & $5.14/2$ & D \\
NGC4278 & 12h20m4.22s & 29\dg16$'$51.24$"$ & $0.00^{+ 0.21}_{- 0.00}$ & $1.34^{+ 0.72}_{- 0.45}$ & $38.38^{+ 0.12}_{- 0.11}$ & $1.79/5$ & D & $0.00^{+ 0.39}_{- 0.00}$ & $3.72^{+ 0.00}_{- 2.43}$ & $38.55^{+ 0.16}_{- 0.21}$ & $0.36/2$ & D \\
NGC4278 & 12h20m5.23s & 29\dg16$'$39.82$"$ & $0.02^{+ 0.28}_{- 0.02}$ & $1.92^{+ 0.82}_{- 0.67}$ & $38.58^{+ 0.09}_{- 0.10}$ & $15.29/8$ & D & $0.00^{+ 0.10}_{- 0.00}$ & $1.64^{+ 0.99}_{- 0.49}$ & $38.54^{+ 0.11}_{- 0.12}$ & $5.18/5$ & D \\
NGC4278 & 12h20m4.33s & 29\dg17$'$35.86$"$ & $0.00^{+ 0.09}_{- 0.00}$ & $1.36^{+ 0.35}_{- 0.28}$ & $38.74^{+ 0.07}_{- 0.07}$ & $17.86/15$ & D & $0.00^{+ 0.20}_{- 0.00}$ & $1.54^{+ 0.72}_{- 0.48}$ & $38.63^{+ 0.10}_{- 0.11}$ & $4.94/6$ & D \\
NGC4278 & 12h20m6.03s & 29\dg16$'$48.25$"$ & $0.02^{+ 0.07}_{- 0.02}$ & $1.45^{+ 0.27}_{- 0.16}$ & $38.95^{+ 0.05}_{- 0.05}$ & $24.28/22$ & P & $0.00^{+ 0.12}_{- 0.00}$ & $1.63^{+ 0.78}_{- 0.44}$ & $38.68^{+ 0.10}_{- 0.10}$ & $8.10/8$ & D \\
NGC4278 & 12h20m5.48s & 29\dg16$'$40.68$"$ & $0.00^{+ 0.07}_{- 0.00}$ & $1.40^{+ 0.32}_{- 0.27}$ & $38.77^{+ 0.06}_{- 0.07}$ & $18.25/18$ & D & $0.00^{+ 0.09}_{- 0.00}$ & $1.83^{+ 1.53}_{- 0.57}$ & $38.75^{+ 0.12}_{- 0.12}$ & $6.50/9$ & D \\
NGC4278 & 12h20m6.79s & 29\dg16$'$56.01$"$ & $0.07^{+ 0.09}_{- 0.07}$ & $1.92^{+ 0.37}_{- 0.33}$ & $38.86^{+ 0.07}_{- 0.06}$ & $35.17/20$ & P & $0.00^{+ 0.06}_{- 0.00}$ & $1.35^{+ 0.36}_{- 0.27}$ & $38.80^{+ 0.07}_{- 0.08}$ & $13.11/14$ & D \\




\#NGC4278 & 12h20m5.95s & 29\dg17$'$8.79$"$ & $0.00^{+ 0.26}_{- 0.00}$ & $1.11^{+ 0.70}_{- 0.24}$ & $38.32^{+ 0.13}_{- 0.16}$ & $3.85/2$ & P & $-$ & $-$ & $<37.93$ & $-$ & $-$ \\ 

NGC1427 & 3h42m18.71s & -35\dg22$'$40.02$"$ & $0.05^{+ 0.11}_{- 0.05}$ & $1.01^{+ 0.33}_{- 0.22}$ & $39.18^{+ 0.07}_{- 0.08}$ & $8.26/10$ & D & $-$ & $-$ & $-$ & $-$ & $-$ \\
NGC1427 & 3h42m18.47s & -35\dg23$'$38.19$"$ & $0.00^{+ 0.04}_{- 0.00}$ & $1.06^{+ 0.36}_{- 0.23}$ & $39.17^{+ 0.09}_{- 0.09}$ & $21.77/11$ & D & $-$ & $-$ & $-$ & $-$ & $-$ \\

  \noalign{\smallskip}\hline
\end{tabular}
\ec
\tablecomments{1.20\textwidth}{\# denotes the sources are present only in first observation. * denotes the sources are present only in second observation. \ddag denotes an additional mekal model added to get better fit for these sources. Host Galaxy Name; Right Ascension; Declination; $n_H$, equivalent hydrogen column density in $10^{22}cm^{-2}$ for the first observation; $\Gamma/kT_{in}$, photon power law index or inner disk temperature in keV in the first observation; $L_{unabs}$, Unabsorbed X-ray luminosity in $\rm ergs~s^{-1}$ in the energy range, 0.3-8.0 keV for the first observation; $\chi^2/d.o.f$, statistics and degree of freedom in the first observation; Best-fit Model (P-Power law, D-Disk black body) in the first observation; $n_H$, equivalent hydrogen column density in $10^{22}cm^{-2}$ for the second observation; $\Gamma/kT_{in}$, photon power law index or inner disk temperature in keV in the second observation; $L_{unabs}$, Unabsorbed X-ray luminosity in $\rm ergs~s^{-1}$ in the energy range, 0.3-8.0 keV for the second observation; $\chi^2/d.o.f$, statistics and degree of freedom in the second observation; Best-fit Model (P-Power law, D-Disk black body) in the second observation; Galactic absorption column density for NGC1399, $n_H$ = $1.53\times10^{20}cm^{-2}$; Galactic absorption column density for NGC4486, $n_H$ = $2.04\times10^{20}cm^{-2}$; Galactic absorption column density for NGC4278, $n_H$ = $1.99\times10^{20}cm^{-2}$; Galactic absorption column density for NGC1427, $n_H$ = $1.63\times10^{20}cm^{-2}$.}
\end{table}

\begin{table}
\bc
\begin{minipage}[]{150mm}
\caption[]{Combined Spectral Properties of point sources fitted with best-fit model\label{xrayvariable}}\end{minipage}
\setlength{\tabcolsep}{3pt}
\small
 \begin{tabular}{ccccccccccc}
  \hline\noalign{\smallskip}
Galaxy & RA (J2000) & Dec (J2000) & $n_H$ & $\Gamma$ & Norm & kT$_{in}$ & Norm & $C_{2}$ & $\chi^2/d.o.f$ & Var(Sig)\\
  \hline\noalign{\smallskip}
NGC1399 & 3h38m32.58s & -35\dg27$'$5.40$"$ & $0.03^{+ 0.03}_{- 0.03}$ & $1.62^{+ 0.10}_{- 0.12}$ & $0.90^{+ 0.09}_{- 0.17}$ & $-$ & $-$ & $1.39^{+ 0.17}_{- 0.15}$ & $58.95/63$ & Y(2.60) \\
NGC1399 & 3h38m31.79s & -35\dg26$'$4.23$"$ & $0.00^{+ 0.02}_{- 0.00}$ & $-$ & $-$ & $0.36^{+ 0.03}_{- 0.03}$ & $0.84^{+ 0.41}_{- 0.26}$ & $1.19^{+ 0.18}_{- 0.15}$ & $33.90/42$ & N(1.27) \\
NGC1399 & 3h38m36.82s & -35\dg27$'$46.98$"$ & $0.00^{+ 0.03}_{- 0.00}$ & $2.03^{+ 0.34}_{- 0.20}$ & $0.20^{+ 0.05}_{- 0.03}$ & $-$ & $-$ & $1.46^{+ 0.45}_{- 0.34}$ & $17.35/15$ & N(1.35) \\
NGC1399 & 3h38m33.09s & -35\dg27$'$31.53$"$ & $0.10^{+ 0.08}_{- 0.08}$ & $2.31^{+ 0.38}_{- 0.31}$ & $0.40^{+ 0.14}_{- 0.18}$ & $-$ & $-$ & $1.41^{+ 0.42}_{- 0.30}$ & $26.02/18$ & N(1.37) \\
NGC1399 & 3h38m25.95s & -35\dg27$'$42.19$"$ & $0.08^{+ 0.27}_{- 0.08}$ & $1.70^{+ 0.48}_{- 0.55}$ & $0.25^{+ 0.27}_{- 0.09}$ & $-$ & $-$ & $0.85^{+ 0.33}_{- 0.24}$ & $ 3.87/8$ & N(0.45) \\
NGC1399 & 3h38m32.76s & -35\dg26$'$58.73$"$ & $0.00^{+ 0.09}_{- 0.00}$ & $1.25^{+ 0.39}_{- 0.32}$ & $0.18^{+ 0.07}_{- 0.04}$ & $-$ & $-$ & $0.90^{+ 0.38}_{- 0.24}$ & $ 9.72/13$ & N(0.26) \\
NGC1399 & 3h38m32.34s & -35\dg27$'$2.11$"$ & $1.64^{+ 1.94}_{- 0.97}$ & $3.42^{+ 2.31}_{- 1.29}$ & $2.25^{+ 26.18}_{- 0.00}$ & $-$ & $-$ & $0.37^{+ 0.30}_{- 0.25}$ & $ 4.77/11$ & Y(2.10) \\
NGC1399 & 3h38m31.86s & -35\dg26$'$49.26$"$ & $0.00^{+ 0.14}_{- 0.00}$ & $2.22^{+ 1.19}_{- 0.37}$ & $0.13^{+ 0.10}_{- 0.13}$ & $-$ & $-$ & $2.32^{+ 2.53}_{- 0.75}$ & $19.20/14$ & N(1.76) \\

NGC4486 & 12h30m47.15s & 12\dg24$'$15.91$"$ & $0.00^{+ 0.01}_{- 0.00}$ & $-$ & $-$ & $0.67^{+ 0.06}_{- 0.06}$ & $0.13^{+ 0.05}_{- 0.04}$ & $1.27^{+ 0.17}_{- 0.15}$ & $177.33/129$ & N(1.80) \\
NGC4486 & 12h30m53.24s & 12\dg23$'$56.69$"$ & $0.23^{+ 0.08}_{- 0.09}$ & $2.27^{+ 0.37}_{- 0.27}$ & $1.40^{+ 0.52}_{- 0.48}$ & $-$ & $-$ & $0.94^{+ 0.26}_{- 0.23}$ & $136.33/108$ & N(0.23) \\
NGC4486 & 12h30m50.12s & 12\dg23$'$1.07$"$ & $0.12^{+ 0.09}_{- 0.12}$ & $2.03^{+ 0.57}_{- 0.43}$ & $0.82^{+ 0.64}_{- 0.34}$ & $-$ & $-$ & $0.70^{+ 0.34}_{- 0.29}$ & $122.94/126$ & N(0.88) \\
NGC4486 & 12h30m46.19s & 12\dg23$'$28.63$"$ & $0.19^{+ 0.07}_{- 0.12}$ & $2.28^{+ 0.50}_{- 0.29}$ & $1.20^{+ 0.70}_{- 0.40}$ & $-$ & $-$ & $1.09^{+ 0.31}_{- 0.26}$ & $112.00/102$ & N(0.35) \\
NGC4486 & 12h30m44.67s & 12\dg22$'$1.06$"$ & $0.23^{+ 0.17}_{- 0.13}$ & $2.41^{+ 0.43}_{- 0.41}$ & $0.89^{+ 0.67}_{- 0.35}$ & $-$ & $-$ & $0.83^{+ 0.36}_{- 0.31}$ & $74.11/71$ & N(0.47) \\
NGC4486 & 12h30m50.80s & 12\dg25$'$2.00$"$ & $0.11^{+ 0.16}_{- 0.11}$ & $1.70^{+ 0.31}_{- 0.36}$ & $0.41^{+ 0.42}_{- 0.10}$ & $-$ & $-$ & $1.40^{+ 0.52}_{- 0.39}$ & $77.62/64$ & N(1.03) \\
NGC4486 & 12h30m44.26s & 12\dg22$'$9.37$"$ & $0.04^{+ 0.22}_{- 0.04}$ & $1.73^{+ 1.10}_{- 0.60}$ & $0.27^{+ 0.29}_{- 0.14}$ & $-$ & $-$ & $1.14^{+ 0.88}_{- 0.57}$ & $90.60/60$ & N(0.25) \\

NGC4278 & 12h20m7.75s & 29\dg17$'$20.39$"$ & $0.08^{+ 0.14}_{- 0.08}$ & $1.44^{+ 0.26}_{- 0.28}$ & $0.29^{+ 0.08}_{- 0.11}$ & $-$ & $-$ & $1.01^{+ 0.22}_{- 0.18}$ & $18.45/19$ & N(0.06) \\
NGC4278 & 12h20m3.43s & 29\dg16$'$39.35$"$ & $0.01^{+ 0.15}_{- 0.01}$ & $-$ & $-$ & $1.38^{+ 1.01}_{- 0.25}$ & $0.13^{+ 0.16}_{- 0.13}$ & $0.69^{+ 0.22}_{- 0.20}$ & $8.59/9$ & N(1.41) \\
NGC4278 & 12h20m4.22s & 29\dg16$'$51.24$"$ & $0.04^{+ 0.28}_{- 0.04}$ & $1.33^{+ 0.50}_{- 0.33}$ & $0.14^{+ 0.12}_{- 0.03}$ & $-$ & $-$ & $1.01^{+ 0.33}_{- 0.27}$ & $3.77/8$ & N(0.04) \\
NGC4278 & 12h20m5.23s & 29\dg16$'$39.82$"$ & $0.07^{+ 0.10}_{- 0.07}$ & $-$ & $-$ & $1.58^{+ 1.01}_{- 0.20}$ & $0.11^{+ 0.08}_{- 0.11}$ & $0.96^{+ 0.24}_{- 0.18}$ & $17.01/14$ & N(0.17) \\
NGC4278 & 12h20m4.33s & 29\dg17$'$35.86$"$ & $0.19^{+ 0.17}_{- 0.11}$ & $1.73^{+ 0.36}_{- 0.24}$ & $0.48^{+ 0.21}_{- 0.15}$ & $-$ & $-$ & $0.69^{+ 0.13}_{- 0.12}$ & $21.98/22$ & Y(2.38) \\
NGC4278 & 12h20m6.03s & 29\dg16$'$48.25$"$ & $0.01^{+ 0.05}_{- 0.01}$ & $1.42^{+ 0.16}_{- 0.16}$ & $0.45^{+ 0.06}_{- 0.10}$ & $-$ & $-$ & $0.64^{+ 0.11}_{- 0.10}$ & $31.79/31$ & Y(3.27) \\
NGC4278 & 12h20m5.48s & 29\dg16$'$40.68$"$ & $0.00^{+ 0.05}_{- 0.00}$ & $-$ & $-$ & $1.50^{+ 0.32}_{- 0.25}$ & $0.22^{+ 0.20}_{- 0.10}$ & $0.79^{+ 0.15}_{- 0.13}$ & $25.50/28$ & N(1.40) \\
NGC4278 & 12h20m6.79s & 29\dg16$'$56.01$"$ & $0.05^{+ 0.03}_{- 0.05}$ & $1.75^{+ 0.14}_{- 0.23}$ & $0.47^{+ 0.06}_{- 0.14}$ & $-$ & $-$ & $1.10^{+ 0.18}_{- 0.16}$ & $47.34/35$ & N(0.63) \\
 \noalign{\smallskip}\hline
\end{tabular}
\ec
\tablecomments{1.17\textwidth}{Host Galaxy Name; Right Ascension; Declination; $n_H$, equivalent hydrogen column density in $10^{22}cm^{-2}$; $\Gamma$, photon power law index; Power law Normalization in $10^{-5}$; $kT_{in}$, inner disk temperature in keV; Disk black body Normalization in $10^{-1}$; $Const_{2}$, Constant2; $\chi^2$ statistics and degree of freedom; X-ray variable (Y-Yes, N-No) and its significance; Galactic absorption column density for NGC1399, $n_H$ = $1.53\times10^{20}cm^{-2}$; Galactic absorption column density for NGC4486, $n_H$ = $2.04\times10^{20}cm^{-2}$; Galactic absorption column density for NGC4278, $n_H$ = $1.99\times10^{20}cm^{-2}$; Constant1, $Const_{1}$ = 1.00.}
\end{table}

\section{Optical counterparts and photometry}
\label{sect:optical}

We search for the optical counterparts for these X-ray sources by using 
the archival {\it HST ACS} images listed in Table \ref{sample}. Typically, the optical sources in 
the {\it HST} images are too faint against the dominant galaxy light and hence to detect 
them, the galaxy light was modelled 
by isophotes of ellipses using the ellipse task in {\sc iraf/stsdas} 
software. The modelled image was then subtracted from the observed galaxy 
image to obtain a residual image. The optical point sources were then extracted 
from the residual image by using {\sc sextractor} with a threshold 
level of $3\sigma$. 

By visual inspection, we could see that for many of the {\it Chandra} X-ray sources within an error 
circle of one arcsecond there is an obvious optical source. However, there was a systematic positional 
offset of one arcsecond between the {\it Chandra} and {\it HST} source positions. This constant positional 
offset was applied to the X-ray sources and then the shifted X-ray positions were compared with the 
optical source positions in  the {\sc sextractor} catalogue. A more detailed explanation with clarifying 
images is presented in \cite{Jit11}. This constant offset is less than the offset of 2.3 arcsec applied 
for the source SN 1993J in the study of a ULX in M81 \citep{Liu02}. We analysed a total of 46 bright 
X-ray sources, which are in the  field of view of {\it HST} images and identified the optical counterpart 
for 34 sources. The optical counterparts identified are unique and for most of the counterparts there is 
no other optical source even within the 3 arcsec from the optical position. The remaining 12 sources 
didn't have an optical counterpart at their respective positions.

Photometry of the optical counterparts as well as all the sources detected by {\sc sextractor} 
was computed on the drizzled images with {\sc iraf/apphot} package. The drizzled images were 
converted from {\it $e^{-}/s/pixel$} to {\it $e^{-}$} per pixel by multiplying the total exposure 
time. An aperture radius of 0.5 arcsec was used to extract the flux by the task {\sc apphot} 
and the magnitudes in the AB magnitude system were calculated using the zero points taken from 
{\it HST ACS} data handbook. The aperture correction were computed from a list of {\sc apphot} photometry files using the {\sc daogrow} algorithm \citep{Ste90} and the correction is applied to the magnitudes. For those X-ray sources that didn't have an optical counterpart 
(i.e. optically dark X-ray sources) we obtained the upper limit of the optical flux at 
the X-ray positions.

\begin{figure}
   \centering
   \includegraphics[width=14.0cm, angle=0]{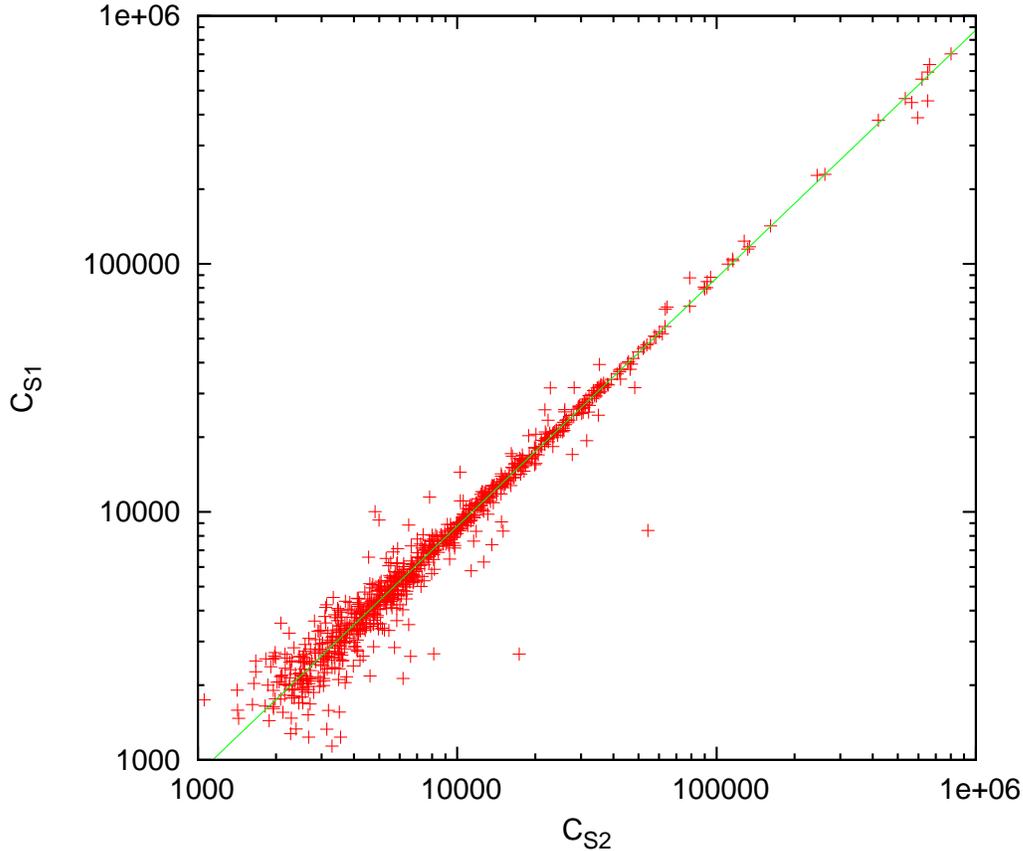}
\caption{The background subtracted counts ($C_{S}$) of the common sources from both observation 
is fitted to a straight-line by the robust estimation method. }
   \label{fluxfit}
   \end{figure}

Our aim is to estimate the optical variability of point sources from two 
observations of a galaxy. This requires a reliable estimate of the statistical and systematic 
errors, if any, in the optical flux. From the photometry, we get the 
total counts, $C$ ($\it{sum}$ from photometry in ADU) and the background subtracted counts, $C_{S}$ 
($\it{flux}$ from photometry in ADU) of each source. The statistical error on $C_{S}$ can be taken to
be $\delta C_{S} = \sqrt{C/epadu}$ where $\it{epadu}$ is the gain parameter in electron per ADU. 
For the two observations of NGC1399, we plot in Figure \ref{fluxfit} the background subtracted 
counts $C_{S1}$ and $C_{S2}$ against each other for 848 sources that are in the common field of view. 
There is the obvious correlation with a large scatter and several outliers. Since there are 
outliers which may affect any least square fitting technique, we use the robust method \citep{Pre92} 
to fit a straight line and obtained a slope $b = 0.876$ and a negligible offset of $a = 5.25$. The two observations have different zero point magnitude ($m_{ZP}$)and exposure time ($T$), which gives this scaling factor (b). For the case of NGC 1399, $m_{1ZP} = 26.059$ and $m_{2ZP} = 26.081$, $T_{1} = 680 sec$ and $T_{2} = 760 sec$ for the two observations. The apparent magnitude, $m = -2.5\times log_{10}(\frac {C_{S}}{T})+2.5\times log_{10}(A)$, where $2.5\times log_{10}(A) = m_{ZP}$. Thus $A = 10^(\frac{m_{ZP}}{2.5})$ and $m = -2.5\times log_{10}(\frac {C_{S}}{T \times A})$. If the apparent magnitudes in the two observations are same, then we can write, $\frac{C_{S2}}{A_{2} \times T_{2}}$ = $\frac{C_{S1}}{A_{1}\times T_{1}}$. Hence $C_{S1} = \frac{A_{1}\times T_{1}}{A_{2}\times T_{2}} \times C_{S2}$. The factor $\frac{A_{1}\times T_{1}}{A_{2}\times T_{2}} = 0.876$ which is as expected identical to the slope $b = 0.876$ obtained by fitting.

\begin{figure}
\centering
\includegraphics[width=14.0cm, angle=0]{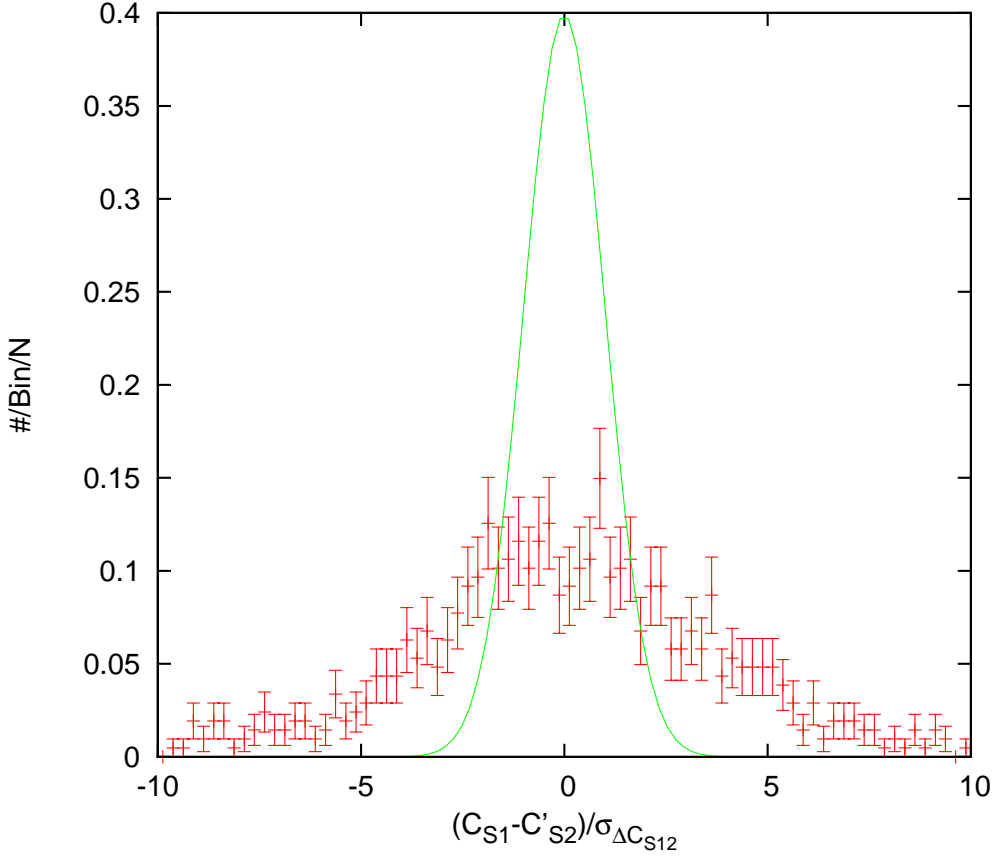}
\caption{The fraction of sources in each bin of $(C_{S1}-C^{\prime}_{S2})/\sigma_{\Delta C_{S12}}$ with 
$\sigma_{\Delta C_{S12}} = \sqrt{\delta {C_{S1}}^2+\delta {C^{\prime}_{S2}}^2}$. The green dotted line is Gaussian 
distribution with mean = 0 and sigma = 1.}
\label{termgauss}
\end{figure}

Then we scaled up the flux of the sources in the second observation i. e., $C^{\prime}_{S2}=b \times C_{S2}+a$ 
and their uncertainties $\delta C^{\prime}_{S2} = b \times \delta C_{S2}$. Now if there were no systematic errors then we
could compare $C_{S1}$ and $C^{\prime}_{S2}$ with their corresponding statistical errors
to determine if a source is variable. However, the statistical errors are small and as
evident in Figure \ref{fluxfit}, this would imply that a large number of the field sources
are variable. Since, we know that this is not the case and indeed most of the field sources
are expected not to vary there is systematic error involved. A better way to illustrate this
is to plot the histogram of $(C_{S1}-C^{\prime}_{S2})/\sigma_{\Delta C_{S12}}$ where 
$\sigma_{\Delta C_{S12}} = \sqrt{\delta {C_{S1}}^2+\delta {C^{\prime}_{S2}}^2}$. If most of the
sources are non-variable and there was no systematic error, then the distribution should
be a zero centred Gaussian with width $\sigma = 1$. However, Figure \ref{termgauss} shows
that as the distribution is significantly broader.

\begin{figure}
\centering
\includegraphics[width=14.0cm, angle=0]{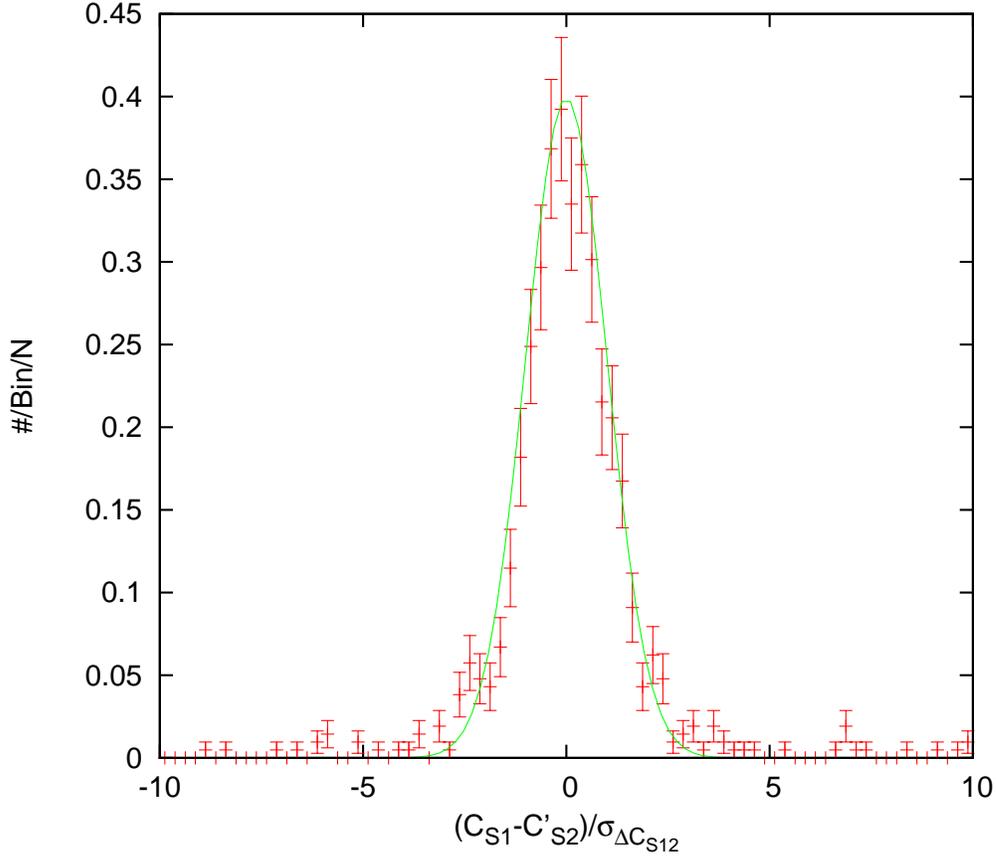}
\caption{The fraction of sources in each bin of $(C_{S1}-C^{\prime}_{S2})/\sigma_{\Delta C_{S12}}$ 
with a systematic (S) added to the uncertainty in flux. The green dotted line is 
Gaussian distribution with mean = 0 and sigma = 1.}
\label{termgaussfit}
\end{figure}

We find that if we add a systematic of $S=525/\sqrt{2}$ to the uncertainties of the flux in quadrature to both
observations, then the distribution is consistent with being a 
Gaussian with $\sigma =1$ as shown in Figure \ref {termgaussfit}. To corroborate that this indeed is
the correct level of systematic error, we do the following exercise. For each pair of optical fluxes, we
compare with a constant and obtain the chi-square, 
\begin{eqnarray}
\chi^2 = \frac{(C_{S1}-C_{S0})^2}{\delta C_{S1}^2} + \frac{(C^{\prime}_{S2}-C_{S0})^2}{\delta {C^{\prime}_{S2}}^2} 
\end{eqnarray}
where $C_{S0}$ is the model constant flux whose value is obtained by minimizing ${\chi}^2$ (i.e. 
$\frac{\partial \chi^2}{\partial C_{S0}} = 0$) to be 
\begin{eqnarray}
C_{S0} = (\frac{C_{S1}}{\delta C_{S1}^2} + \frac{C^{\prime}_{S2}}{\delta {C^{\prime}_{S2}}^2}) (\frac{\delta C_{S1}^2 \delta {C^{\prime}_{S2}}^2}{\delta C_{S1}^2 + \delta {C^{\prime}_{S2}}^2})
\end{eqnarray}
Since the number of data points is two and the number of parameters (i.e. $C_{S0}$) is one, the degree of
freedom here is one. Hence, if the model for a majority of the sources (i.e. the sources
are not variable) and the error estimates are correct then the distribution of $\chi^2$ should
be a chi-square distribution of order one i.e.
\begin{eqnarray}
P(x) = \frac{1}{\sqrt{2\pi}} x^{-1/2} \exp(-x/2) 
\end{eqnarray}

\begin{figure}
\centering
\includegraphics[width=14.0cm, angle=0]{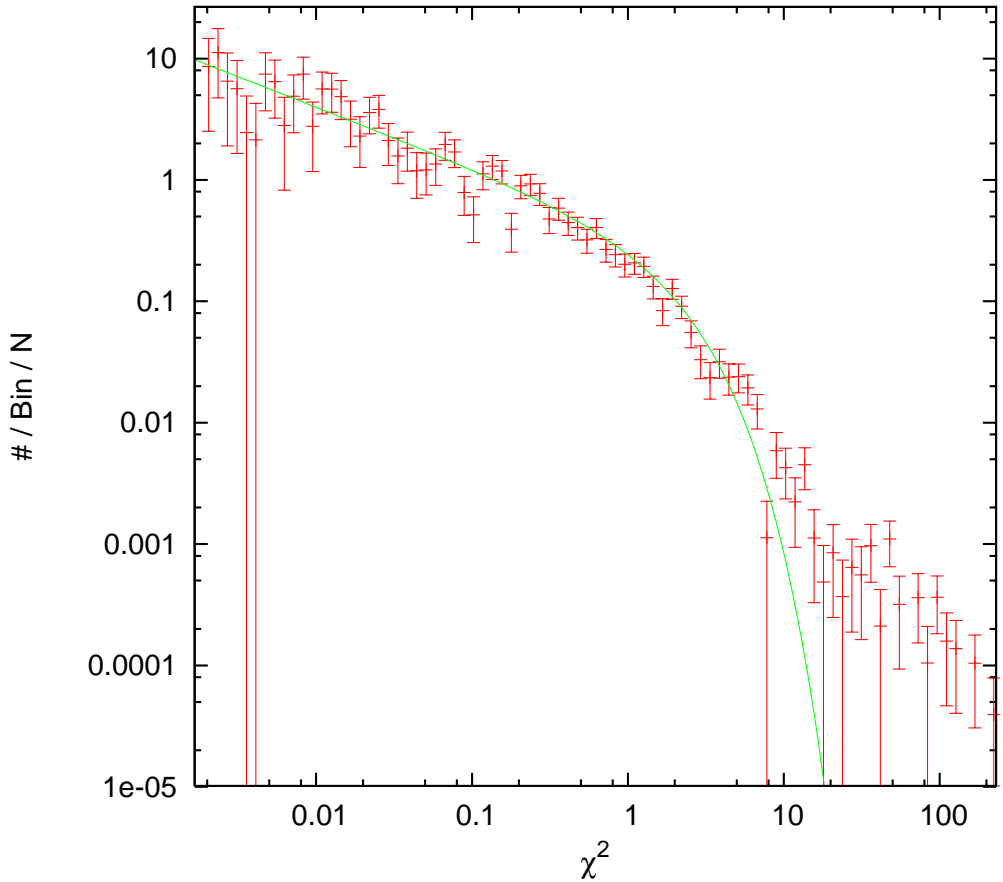}
\caption{The fraction of sources in each bin of ${\chi}^2$. The green dotted line is 
chi-square distribution, P(x).}
\label{chidistribution}
\end{figure}

Figure \ref {chidistribution} shows the distribution of $\chi^2$ for all the 848 sources in NGC1399. The
solid line is the expected distribution $P(x)$. For a majority of the sources which are expected not
to be variable $\chi^2 < 2$ as expected. More importantly the distribution matches well with the
majority including the low $\chi^2$ values of $\sim 0.01$. This strongly implies that the
systematic error used is reliable. 
We could not identify the cause for the systematic errors despite our best efforts. However, we note that such deviations have been reported in similar works. For example, for NGC 1313, \cite{Liu07} reported that out of 399 optical sources they examined, more than 81 (i.e. 20\%) had variability above 2 sigma, while the expected number was more like 10\%. 
The measured distribution deviates from the expected one
for $\chi^2 > 12$ and these are the few truly variable sources in the sample. Thus we can
state confidently and conservatively that sources with $\chi^2 > 12$ are indeed variable and we use this criterion
for this work. About 93\% (792) cross identified sources are not variable between the two 
observations and 56 sources are optically variable i.e ${\chi}^2 {\geq}12$. 
While the results presented above are for NGC1399, we use the same technique to establish the systematic
error for the other three galaxies and for each of them we find that $\chi^2 > 12$ to be a good conservative
criterion for optical variability. The photometric optical magnitudes of the X-ray sources of the sample
have been provided in Table \ref{optvariable}.

Table \ref{IR} provides the properties of the four X-ray sources which are 
optically variable. We have also estimated the difference in the magnitude of these sources by comparing the 
F814W and F850LP data. Even though they are different bands, 
three sources (source 2 and source 3 in NGC1399, one source in NGC1427) 
show a magnitude difference of 0.1 - 0.4. But the Source 1 in NGC1399 has a magnitude difference of 0.02 only in these filters.

\begin{table}
\bc
\begin{minipage}[]{150mm}
\caption[]{The Variability of Optical Counterparts in the Sample\label{optvariable}}\end{minipage}
\setlength{\tabcolsep}{4.0pt}
\scriptsize
 \begin{tabular}{ccccccccccc}
  \hline\noalign{\smallskip}
Galaxy & RA (J2000) & Dec (J2000) & log($L_{1}$) & Model & $C_{2}$ & Var(Sig) & $m_{1}$ & $m_{2}$ & ${\Delta m}$ & ${\chi}^2$\\
  \hline\noalign{\smallskip}

NGC1399 & 3h38m32.58s & -35\dg27$'$5.40$"$ & $39.39^{+ 0.06}_{- 0.05}$ & P & $1.39^{+ 0.17}_{- 0.15}$ & Y(2.60) & $22.157\pm0.018$ & $22.169\pm0.019$ & $-0.012\pm0.026$ & 0.259 \\
NGC1399 & 3h38m31.79s & -35\dg26$'$4.23$"$ & $39.04^{+ 0.15}_{- 0.06}$ & D & $1.19^{+ 0.18}_{- 0.15}$ & N(1.27) & $22.787\pm0.031$ & $22.779\pm0.031$ & $0.008\pm0.044$ & 0.031 \\
NGC1399 & 3h38m36.82s & -35\dg27$'$46.98$"$ & $38.72^{+ 0.10}_{- 0.11}$ & P & $1.46^{+ 0.45}_{- 0.34}$ & N(1.35) & $23.231\pm0.045$ & $22.847\pm0.032$ & $0.384\pm0.055$ & 51.437 \\
NGC1399 & 3h38m33.09s & -35\dg27$'$31.53$"$ & $38.61^{+ 0.12}_{- 0.08}$ & D & $1.41^{+ 0.42}_{- 0.30}$ & N(1.37) & $21.692\pm0.012$ & $21.607\pm0.011$ & $0.085\pm0.016$ & 26.493 \\
NGC1399 & 3h38m33.09s & -35\dg27$'$31.53$"$ & $38.61^{+ 0.12}_{- 0.08}$ & D & $1.41^{+ 0.42}_{- 0.30}$ & N(1.37) & $23.188\pm0.044$ & $23.110\pm0.042$ & $0.078\pm0.061$ & 1.575 \\
NGC1399 & 3h38m25.95s & -35\dg27$'$42.19$"$ & $38.64^{+ 0.18}_{- 0.15}$ & D & $0.85^{+ 0.33}_{- 0.24}$ & N(0.45) & $ > 26.646$ & $ > 26.656$ & $-$ & $-$\\
NGC1399 & 3h38m32.76s & -35\dg26$'$58.73$"$ & $38.86^{+ 0.19}_{- 0.19}$ & D & $0.90^{+ 0.38}_{- 0.24}$ & N(0.26) & $23.070\pm0.040$ & $23.040\pm0.041$ & $0.030\pm0.058$ & 0.260 \\
NGC1399 & 3h38m32.34s & -35\dg27$'$2.11$"$ & $38.94^{+ 0.52}_{- 0.25}$ & D & $0.37^{+ 0.30}_{- 0.25}$ & Y(2.10) & $24.446\pm0.143$ & $24.525\pm0.160$ & $-0.079\pm0.215$ & 0.140 \\
NGC1399 & 3h38m31.86s & -35\dg26$'$49.26$"$ & $38.41^{+ 0.58}_{- 0.31}$ & D & $2.32^{+ 2.53}_{- 0.75}$ & N(1.76) & $21.878\pm0.014$ & $22.146\pm0.019$ & $-0.268\pm0.024$ & 133.152 \\
NGC1399 & 3h38m25.66s & -35\dg27$'$41.50$"$ & $38.67^{+ 0.37}_{- 0.18}$ & D & $-$ & $-$ & $24.860\pm0.203$ & $25.233\pm0.291$ & $-0.373\pm0.355$ & 1.197 \\
NGC1399 & 3h38m27.80s & -35\dg25$'$26.65$"$ & $38.55^{+ 0.17}_{- 0.18}$ & D & $-$ & $-$ & $24.783\pm0.186$ & $24.820\pm0.192$ & $-0.037\pm0.267$ & 0.019 \\
NGC1399 & 3h38m26.50s & -35\dg27$'$32.29$"$ & $ < 38.08$ & $-$ & $-$ & $-$ & $ > 26.661$ & $ > 26.697$ & $-$ & $-$\\
NGC1399 & 3h38m33.82s & -35\dg25$'$56.95$"$ & $ < 37.94$ & $-$ & $-$ & $-$ & $20.870\pm0.006$ & $20.866\pm0.006$ & $0.004\pm0.008$ & 0.062 \\
NGC1399 & 3h38m33.80s & -35\dg26$'$58.30$"$ & $ < 38.73$ & $-$ & $-$ & $-$ & $22.832\pm0.032$ & $22.849\pm0.034$ & $-0.017\pm0.046$ & 0.159 \\
NGC1399 & 3h38m32.35s & -35\dg27$'$10.63$"$ & $ < 38.16$ & $-$ & $-$ & $-$ & $23.046\pm0.040$ & $23.084\pm0.043$ & $-0.038\pm0.059$ & 0.444 \\
NGC1399 & 3h38m25.32s & -35\dg27$'$53.49$"$ & $ < 38.20$ & $-$ & $-$ & $-$ & $22.256\pm0.019$ & $22.257\pm0.020$ & $-0.001\pm0.027$ & 0.003 \\
NGC1399 & 3h38m27.19s & -35\dg26$'$1.53$"$ & $ < 38.38$ & $-$ & $-$ & $-$ & $22.206\pm0.018$ & $22.197\pm0.019$ & $0.009\pm0.026$ & 0.080 \\
NGC1399 & 3h38m27.63s & -35\dg26$'$48.54$"$ & $39.23^{+ 0.23}_{- 0.15}$ & P & $1.14^{+ 0.21}_{- 0.16}$ & N(0.87) & $ > 25.979$ & $ > 25.894$ & $-$ & $-$\\
NGC1399 & 3h38m38.76s & -35\dg25$'$54.86$"$ & $39.09^{+ 0.09}_{- 0.18}$ & D & $0.22^{+ 0.12}_{- 0.11}$ & Y(6.50) & $21.740\pm0.012$ & $21.732\pm0.012$ & $0.008\pm0.017$ & 0.165 \\

NGC4486 & 12h30m47.15s & 12\dg24$'$15.91$"$ & $39.17^{+ 0.04}_{- 0.04}$ & D & $1.27^{+ 0.17}_{- 0.15}$ & N(1.80) & $22.416\pm0.044$ & $22.512\pm0.049$ & $-0.096\pm0.066$ & 2.409 \\
NGC4486 & 12h30m47.15s & 12\dg24$'$15.91$"$ & $39.17^{+ 0.04}_{- 0.04}$ & D & $1.27^{+ 0.17}_{- 0.15}$ & N(1.80) & $22.786\pm0.062$ & $22.744\pm0.060$ & $0.042\pm0.087$ & 0.161 \\
NGC4486 & 12h30m53.24s & 12\dg23$'$56.69$"$ & $39.03^{+ 0.07}_{- 0.08}$ & D & $0.94^{+ 0.26}_{- 0.23}$ & N(0.23) & $22.901\pm0.069$ & $22.826\pm0.065$ & $0.075\pm0.094$ & 0.520 \\
NGC4486 & 12h30m50.12s & 12\dg23$'$1.07$"$ & $38.97^{+ 0.08}_{- 0.10}$ & D & $0.70^{+ 0.34}_{- 0.29}$ & N(0.88) & $20.686\pm0.010$ & $20.668\pm0.010$ & $0.018\pm0.014$ & 1.354 \\
NGC4486 & 12h30m46.19s & 12\dg23$'$28.63$"$ & $38.95^{+ 0.07}_{- 0.08}$ & D & $1.09^{+ 0.31}_{- 0.26}$ & N(0.35) & $20.408\pm0.007$ & $20.417\pm0.007$ & $-0.009\pm0.010$ & 1.261 \\
NGC4486 & 12h30m44.67s & 12\dg22$'$1.06$"$ & $39.16^{+ 0.47}_{- 0.21}$ & P & $0.83^{+ 0.36}_{- 0.31}$ & N(0.47) & $21.471\pm0.019$ & $21.470\pm0.019$ & $0.001\pm0.026$ & 0.008 \\

NGC4486 & 12h30m50.80s & 12\dg25$'$2.00$"$ & $38.82^{+ 0.10}_{- 0.12}$ & D & $1.40^{+ 0.52}_{- 0.39}$ & N(1.03) & $ > 26.367$ & $ > 26.260$ & $-$ & $-$ \\
NGC4486 & 12h30m44.26s & 12\dg22$'$9.37$"$ & $38.36^{+ 0.46}_{- 0.19}$ & D & $1.14^{+ 0.88}_{- 0.57}$ & N(0.25) & $ > 26.478$ & $ > 26.333$ & $-$ & $-$ \\
 
NGC4486 & 12h30m44.71s & 12\dg24$'$34.61$"$ & $39.08^{+ 0.06}_{- 0.05}$ & P & $-$ & $-$ & $21.800\pm0.025$ & $21.763\pm0.024$ & $0.037\pm0.035$ & 0.887 \\
NGC4486 & 12h30m46.32s & 12\dg23$'$23.19$"$ & $38.89^{+ 0.11}_{- 0.08}$ & D & $-$ & $-$ & $20.364\pm0.007$ & $20.351\pm0.007$ & $0.013\pm0.010$ & 1.278 \\
NGC4486 & 12h30m47.32s & 12\dg23$'$8.82$"$ & $38.84^{+ 0.15}_{- 0.11}$ & D & $-$ & $-$ & $20.742\pm0.010$ & $20.762\pm0.010$ & $-0.020\pm0.014$ & 2.599 \\
NGC4486 & 12h30m50.08s & 12\dg22$'$51.21$"$ & $38.46^{+ 0.13}_{- 0.27}$ & D & $-$ & $-$ & $20.902\pm0.011$ & $20.904\pm0.012$ & $-0.002\pm0.016$ & 0.094 \\
NGC4486 & 12h30m52.79s & 12\dg23$'$36.85$"$ & $44.10^{+ 5.62}_{- 2.00}$ & P & $-$ & $-$ & $23.332\pm0.103$ & $23.523\pm0.125$ & $-0.191\pm0.162$ & 1.617 \\
NGC4486 & 12h30m43.49s & 12\dg23$'$46.80$"$ & $38.34^{+ 0.74}_{- 0.28}$ & D & $-$ & $-$ & $20.067\pm0.005$ & $20.076\pm0.005$ & $-0.009\pm0.007$ & 2.220 \\
NGC4486 & 12h30m46.52s & 12\dg24$'$50.15$"$ & $38.40^{+ 0.50}_{- 0.19}$ & D & $-$ & $-$ & $23.436\pm0.111$ & $23.373\pm0.106$ & $0.063\pm0.154$ & 0.111 \\

NGC4486 & 12h30m44.91s & 12\dg24$'$4.50$"$ & $38.43^{+ 0.19}_{- 0.27}$ & D & $-$ & $-$ & $ > 26.199$ & $ > 26.086$ & $-$ & $-$ \\
NGC4486 & 12h30m50.82s & 12\dg24$'$11.80$"$ & $38.80^{+ 0.19}_{- 0.15}$ & D & $-$ & $-$ & $ > 25.733$ & $ > 25.873$ & $-$ & $-$ \\
NGC4486 & 12h30m49.13s & 12\dg21$'$59.40$"$ & $38.69^{+ 17.17}_{- 9.11}$ & P & $-$ & $-$ & $ > 26.154$ & $ > 26.032$ & $-$ & $-$ \\

NGC4278 & 12h20m7.75s & 29\dg17$'$20.39$"$ & $38.64^{+ 0.08}_{- 0.09}$ & D & $1.01^{+ 0.22}_{- 0.18}$ & N(0.06) & $20.283\pm0.010$ & $20.294\pm0.010$ & $-0.011\pm0.013$ & 0.001 \\
NGC4278 & 12h20m3.43s & 29\dg16$'$39.35$"$ & $38.49^{+ 0.13}_{- 0.13}$ & D & $0.69^{+ 0.22}_{- 0.20}$ & N(1.41) & $21.205\pm0.021$ & $21.188\pm0.021$ & $0.017\pm0.029$ & 0.823 \\ 
NGC4278 & 12h20m4.22s & 29\dg16$'$51.24$"$ & $38.38^{+ 0.12}_{- 0.11}$ & D & $1.01^{+ 0.33}_{- 0.27}$ & N(0.04) & $21.334\pm0.024$ & $21.374\pm0.025$ & $-0.040\pm0.034$ & 0.864 \\ 
NGC4278 & 12h20m5.23s & 29\dg16$'$39.82$"$ & $38.58^{+ 0.09}_{- 0.10}$ & D & $0.96^{+ 0.24}_{- 0.18}$ & N(0.17) & $20.998\pm0.018$ & $21.017\pm0.019$ & $-0.019\pm0.026$ & 0.111 \\




NGC4278 & 12h20m4.33s & 29\dg17$'$35.86$"$ & $38.74^{+ 0.07}_{- 0.07}$ & D & $0.69^{+ 0.13}_{- 0.12}$ & Y(2.38) & $ > 25.952$ & $ > 25.916$ & $-$ & $-$ \\
NGC4278 & 12h20m6.03s & 29\dg16$'$48.25$"$ & $38.95^{+ 0.05}_{- 0.05}$ & P & $0.64^{+ 0.11}_{- 0.10}$ & Y(3.27) & $ > 24.455$ & $ > 24.497$ & $-$ & $-$ \\
NGC4278 & 12h20m5.48s & 29\dg16$'$40.68$"$ & $38.77^{+ 0.06}_{- 0.07}$ & D & $0.79^{+ 0.15}_{- 0.13}$ & N(1.40) & $ > 24.990$ & $ > 25.018$ & $-$ & $-$ \\
NGC4278 & 12h20m6.79s & 29\dg16$'$56.01$"$ & $38.86^{+ 0.07}_{- 0.06}$ & P & $1.10^{+ 0.18}_{- 0.16}$ & N(0.63) & $ > 23.504$ & $ > 23.493$ & $-$ & $-$ \\

NGC4278 & 12h20m5.95s & 29\dg17$'$8.79$"$ & $38.32^{+ 0.13}_{- 0.16}$ & P & $-$ & $-$ & $22.014\pm0.047$ & $21.945\pm0.043$ & $0.069\pm0.064$ & 1.388 \\

NGC1427 & 3h42m18.71s & -35\dg22$'$40.02$"$ & $39.18^{+0.07}_{-0.08}$ & D & $-$ & $-$ & $22.450\pm0.022$ & $22.461\pm0.022$ & $-0.011\pm0.031$ & 0.163 \\
NGC1427 & 3h42m18.47s & -35\dg23$'$38.19$"$ & $39.17^{+0.09}_{-0.09}$ & D & $-$ & $-$ & $22.861\pm0.036$ & $23.147\pm0.052$ & $-0.286\pm0.063$ & 19.488 \\

 \noalign{\smallskip}\hline
\end{tabular}
\ec
\tablecomments{1.15\textwidth}{(1) Host Galaxy Name; (2) Right Ascension; (3) Declination; (4) Log of unabsorbed X-ray luminosity in $\rm ergs~s^{-1}$ for first observation; (5) Best-fit Model; (6) Constant2; (7) X-ray variable (Y-Yes, N-No) and its significance; (8) Aperture corrected magnitude in the first observation; (9) Aperture corrected magnitude in the second observation; (10) The difference in magnitude; (11) Significance of the Optical variability. In the sample, two sources (3h38m33.09s, -35\dg27$'$31.53$"$ in NGC1399 and 12h30m47.15s, 12\dg24$'$15.91$"$ in NGC4486) have two possible optical counterparts. Hence we report the magnitude of each counterpart.}
\end{table}

\begin{table}
\bc
\begin{minipage}[]{150mm}
\caption[]{The Properties of Optically Varying Sources in the Sample\label{IR}}\end{minipage}
\setlength{\tabcolsep}{2.5pt}
\small
 \begin{tabular}{cccccccccc}
  \hline\noalign{\smallskip}
Galaxy & RA (J2000) & Dec (J2000) & log($L_{1}$) & ${\Delta m}$ & ${\chi}^2$ & $g-z$ & $F_{3.6\mu m}$ & $F_{5.8\mu m}$ & $F_{5.8}/F_{3.6}$\\
  \hline\noalign{\smallskip}
NGC1399 & 3h38m31.86s & -35\dg26$'$49.26$"$ & $38.41^{+ 0.58}_{- 0.31}$ & $-0.268\pm0.024$ & $133.152$ & $1.048$ & $42.37\pm0.83$ & $25.00\pm1.98$ & $0.59$ \\
NGC1399 & 3h38m33.09s & -35\dg27$'$31.53$"$ & $38.61^{+ 0.12}_{- 0.08}$ & $0.085\pm0.016$ & $26.493$ & $1.307$ & $24.58\pm0.59$ & $32.19\pm2.10$ & $1.31$ \\
NGC1399 & 3h38m36.82s & -35\dg27$'$46.98$"$ & $38.72^{+ 0.10}_{- 0.11}$ & $0.384\pm0.055$ & $51.437$ & $1.842$ & $7.41\pm0.42$ & $< 5.38$ & $< 0.73$ \\
NGC1427 & 3h42m18.47s & -35\dg23$'$38.19$"$ & $39.17^{+ 0.09}_{- 0.09}$ & $-0.286\pm0.063$ & $19.488$ & $1.835$ & $<2.78$ & $<3.93$ & $-$ \\
 \noalign{\smallskip}\hline
\end{tabular}
\ec
\tablecomments{1.12\textwidth}{(1) Host Galaxy Name; (2) Right Ascension; (3) Declination; (4) log of unabsorbed X-ray luminosity in $\rm ergs~s^{-1}$ for first observation; (5) The difference in magnitude; (6) Significance of the Optical variability; (7) Optical colour (g-z) derived from Vega magnitude; (8),(9) IR flux in mJy for the 3.6$\mu$m and 5.8$\mu$m bands; (10) Mid-IR flux ratio.}
\end{table}

\section{Discussion}
\label{sect:discussion}

In this work we have studied the long term X-ray and optical variability of X-ray sources in four nearby elliptical 
galaxies. For the 46 sources in the sample, we have fitted their X-ray spectra using an absorbed power-law or black body model
for two {\it Chandra} observations and found that 24 of them show long term X-ray variability. For 34 sources,
we have identified optical counterparts. After estimating the systematic error on the photometric magnitude, we find
that four of the sources clearly exhibit long term optical variation. Since the optical counterpart is varying 
it cannot be the integrated light of stars in a globular cluster. Thus, one
may expect that the optical variability is induced by the X-ray source. If that is so, these sources are important candidates
for further study.

The optically variable X-ray sources could be background Active Galactic Nuclei (AGN). 
The reported optical colours ({\it g - z}) for the
sources in NGC1399 \citep{Sha13} are tabulated in Table \ref{IR} and they reveal that the objects are blue and one of them is
bluer than blue globular clusters, $1.3 < g-z < 1.9$ \citep{Pao11}. Indeed, the optically variable sources 
(Source 1 and 2 in NGC1399) were identified as possible contaminants in an earlier analysis 
\citep{Kun07}. The analysis of {\it HST/WFPC} data reveals that these sources are bluer than ${\it B-I=1.5}$ and hence are not 
globular clusters. \cite {Bla12} studied the globular cluster systems in NGC1399 using the {\it HST/ACS g, V, I, z} and {\it H} bands. 
In their study, the sources with $19.5 < I_{814} < 23.5$ and $0.5 < g_{475} - I_{814} < 1.6$ are classified as the globular clusters.
and the optically variable sources in NGC1399 (source 1 and 2) again do not satisfy their criteria. This may indicate that they
may be background AGN and indeed their IR colours also support this interpretation. Studies have shown that AGN have 
flux ratios $> 0.63$ in the $5.8$ and $3.6 \mu m$ bands i.e. $F_{5.8}/F_{3.6} > 0.63$ \citep{Pol06,Lac04}. 
\citet{Sha13} have looked for IR counterparts of X-ray sources in NGC1399 using {\it Spitzer} data. Their
quoted IR flux and ratios are tabulated in Table \ref{IR}. All four sources have IR flux ratios $ \geq 0.63$, indicating that 
they maybe background AGN. Unfortunately these sources are not in the field of view 
of the {\it Spitzer} $4.5$ and $8.0 \mu m$ images, which would have provided more information on the nature
of these sources.

We do not find evidence for any optical counterpart to disappear or flux changes by order of magnitude. Such variations
would be expected if the X-ray emission is due to a violent transient event like a very bright nova explosion or
a tidal disruption of a white dwarf by a black hole. Such transient events are expected to show dramatic variation in both
X-ray and optical flux. While there are several X-ray sources which are not detected in the other {\it Chandra} observation,
none of them exhibit dramatic variability in the optical. For example, as mentioned earlier, \citet{Irw10} have argued
that the lack of H$\alpha$ and H$\beta$ in the spectrum of a ULXs in NGC1399 (CXOJ033831.8-352604) may indicate the
tidal disruption of a white dwarf by a black hole. However, here we find that neither the X-ray nor the optical flux
show any long term variation.

Clearly, conclusive evidence on the nature of these sources can be obtained only by studying their
optical spectra and confirming by emission line studies whether a source is a background AGN or not. Such studies will also
provide clear information about the origin of
the optical source. Since this would require large telescopes in excellent seeing conditions, it is important to
choose good potential candidates such as the optically variable sources identified here. A positive identification of
a optically variable source as not being a background AGN, 
would be the crucial step towards understanding these enigmatic sources.

\normalem
\begin{acknowledgements}

VJ, KJ, CDR, and BRSB thank the IUCAA visitors program and UGC Special assistance 
program. VJ acknowledges financial support from the Council of Scientific and Industrial Research 
(CSIR) through SRF scheme. This work has been partially funded from the ISRO-RESPOND program. PS 
would like to thank the DST - FAST Track Scheme for research funding. The authors thank Phil 
Charles for useful discussions.

\end{acknowledgements}



\bibliographystyle{raa}
\bibliography{ulx}

\begin{thebibliography}{37}
\providecommand{\natexlab}[1]{#1}
\providecommand{\selectlanguage}[1]{\relax}

\bibitem[{{Angelini} et~al.(2001){Angelini}, {Loewenstein}, \&
  {Mushotzky}}]{Ang01}
{Angelini}, L., {Loewenstein}, M., \& {Mushotzky}, R.~F. 2001, \apjl, 557, L35

\bibitem[{{Bassino} et~al.(2006){Bassino}, {Faifer}, {Forte} et~al.}]{Bas06}
{Bassino}, L.~P., {Faifer}, F.~R., {Forte}, J.~C., et~al. 2006, \aap, 451, 789

\bibitem[{{Blakeslee} et~al.(2012){Blakeslee}, {Cho}, {Peng} et~al.}]{Bla12}
{Blakeslee}, J.~P., {Cho}, H., {Peng}, E.~W., et~al. 2012, \apj, 746, 88

\bibitem[{{Brassington} et~al.(2009){Brassington}, {Fabbiano}, {Kim}
  et~al.}]{Bra09}
{Brassington}, N.~J., {Fabbiano}, G., {Kim}, D.-W., et~al. 2009, \apjs, 181,
  605

\bibitem[{{Dirsch} et~al.(2003){Dirsch}, {Richtler}, {Geisler} et~al.}]{Dir03}
{Dirsch}, B., {Richtler}, T., {Geisler}, D., et~al. 2003, \aj, 125, 1908

\bibitem[{{Fabbiano}(1989)}]{Fab89}
{Fabbiano}, G. 1989, \araa, 27, 87

\bibitem[{{Fabbiano} et~al.(2010){Fabbiano}, {Brassington}, {Lentati}
  et~al.}]{Fab10}
{Fabbiano}, G., {Brassington}, N.~J., {Lentati}, L., et~al. 2010, \apj, 725,
  1824

\bibitem[{{Forte} et~al.(2001){Forte}, {Geisler}, {Ostrov}, {Piatti}, \&
  {Gieren}}]{For01}
{Forte}, J.~C., {Geisler}, D., {Ostrov}, P.~G., {Piatti}, A.~E., \& {Gieren},
  W. 2001, \aj, 121, 1992

\bibitem[{{Goad} et~al.(2002){Goad}, {Roberts}, {Knigge}, \& {Lira}}]{Goa02}
{Goad}, M.~R., {Roberts}, T.~P., {Knigge}, C., \& {Lira}, P. 2002, \mnras, 335,
  L67

\bibitem[{{Gris{\'e}} et~al.(2011){Gris{\'e}}, {Kaaret}, {Pakull}, \&
  {Motch}}]{Gri11}
{Gris{\'e}}, F., {Kaaret}, P., {Pakull}, M.~W., \& {Motch}, C. 2011, \apj, 734,
  23

\bibitem[{{Gris{\'e}} et~al.(2008){Gris{\'e}}, {Pakull}, {Soria}
  et~al.}]{Gri08}
{Gris{\'e}}, F., {Pakull}, M.~W., {Soria}, R., et~al. 2008, \aap, 486, 151

\bibitem[{{Harris}(1991)}]{Har91}
{Harris}, W.~E. 1991, \araa, 29, 543

\bibitem[{{Irwin}(2006)}]{Irw06}
{Irwin}, J.~A. 2006, \mnras, 371, 1903

\bibitem[{{Irwin} et~al.(2010){Irwin}, {Brink}, {Bregman}, \&
  {Roberts}}]{Irw10}
{Irwin}, J.~A., {Brink}, T.~G., {Bregman}, J.~N., \& {Roberts}, T.~P. 2010,
  \apjl, 712, L1

\bibitem[{{Jithesh} et~al.(2011){Jithesh}, {Jeena}, {Misra} et~al.}]{Jit11}
{Jithesh}, V., {Jeena}, K., {Misra}, R., et~al. 2011, \apj, 729, 67

\bibitem[{{Jord{\'a}n} et~al.(2004){Jord{\'a}n}, {C{\^o}t{\'e}}, {Ferrarese}
  et~al.}]{Jor04}
{Jord{\'a}n}, A., {C{\^o}t{\'e}}, P., {Ferrarese}, L., et~al. 2004, \apj, 613,
  279

\bibitem[{{Kaaret} et~al.(2001){Kaaret}, {Prestwich}, {Zezas} et~al.}]{Kaa01}
{Kaaret}, P., {Prestwich}, A.~H., {Zezas}, A., et~al. 2001, \mnras, 321, L29

\bibitem[{{Kim} et~al.(2009){Kim}, {Fabbiano}, {Brassington} et~al.}]{Kim09}
{Kim}, D., {Fabbiano}, G., {Brassington}, N.~J., et~al. 2009, \apj, 703, 829

\bibitem[{{Kim} et~al.(2006){Kim}, {Kim}, {Fabbiano} et~al.}]{Kim06}
{Kim}, E., {Kim}, D., {Fabbiano}, G., et~al. 2006, \apj, 647, 276

\bibitem[{{Kissler-Patig} et~al.(1997){Kissler-Patig}, {Kohle}, {Hilker}
  et~al.}]{Kis97}
{Kissler-Patig}, M., {Kohle}, S., {Hilker}, M., et~al. 1997, \aap, 319, 470

\bibitem[{{Kong} et~al.(2006){Kong}, {Charles}, {Homer}, {Kuulkers}, \&
  {O'Donoghue}}]{Kon06}
{Kong}, A.~K.~H., {Charles}, P.~A., {Homer}, L., {Kuulkers}, E., \&
  {O'Donoghue}, D. 2006, \mnras, 368, 781

\bibitem[{{Kundu} et~al.(2007){Kundu}, {Maccarone}, \& {Zepf}}]{Kun07}
{Kundu}, A., {Maccarone}, T.~J., \& {Zepf}, S.~E. 2007, \apj, 662, 525

\bibitem[{{Lacy} et~al.(2004){Lacy}, {Storrie-Lombardi}, {Sajina}
  et~al.}]{Lac04}
{Lacy}, M., {Storrie-Lombardi}, L.~J., {Sajina}, A., et~al. 2004, \apjs, 154,
  166

\bibitem[{{Liu}(2011)}]{Liu11}
{Liu}, J. 2011, \apjs, 192, 10

\bibitem[{{Liu} et~al.(2007){Liu}, {Bregman}, {Miller}, \& {Kaaret}}]{Liu07}
{Liu}, J., {Bregman}, J., {Miller}, J., \& {Kaaret}, P. 2007, \apj, 661, 165

\bibitem[{{Liu} et~al.(2002){Liu}, {Bregman}, \& {Seitzer}}]{Liu02}
{Liu}, J., {Bregman}, J.~N., \& {Seitzer}, P. 2002, \apjl, 580, L31

\bibitem[{{Makishima} et~al.(2000){Makishima}, {Kubota}, {Mizuno}
  et~al.}]{Mak00}
{Makishima}, K., {Kubota}, A., {Mizuno}, T., et~al. 2000, \apj, 535, 632

\bibitem[{{Mucciarelli} et~al.(2007){Mucciarelli}, {Zampieri}, {Treves},
  {Turolla}, \& {Falomo}}]{Muc07}
{Mucciarelli}, P., {Zampieri}, L., {Treves}, A., {Turolla}, R., \& {Falomo}, R.
  2007, \apj, 658, 999

\bibitem[{{Paolillo} et~al.(2011){Paolillo}, {Puzia}, {Goudfrooij}
  et~al.}]{Pao11}
{Paolillo}, M., {Puzia}, T.~H., {Goudfrooij}, P., et~al. 2011, \apj, 736, 90

\bibitem[{{Polletta} et~al.(2006){Polletta}, {Wilkes}, {Siana} et~al.}]{Pol06}
{Polletta}, M.~d.~C., {Wilkes}, B.~J., {Siana}, B., et~al. 2006, \apj, 642, 673

\bibitem[{{Press} et~al.(1992){Press}, {Teukolsky}, {Vetterling}, \&
  {Flannery}}]{Pre92}
{Press}, W.~H., {Teukolsky}, S.~A., {Vetterling}, W.~T., \& {Flannery}, B.~P.
  1992, {Numerical recipes in FORTRAN. The art of scientific computing}

\bibitem[{{Ptak} et~al.(2006){Ptak}, {Colbert}, {van der Marel} et~al.}]{Pta06}
{Ptak}, A., {Colbert}, E., {van der Marel}, R.~P., et~al. 2006, \apjs, 166, 154

\bibitem[{{Shalima} et~al.(2013){Shalima}, {Jithesh}, {Jeena} et~al.}]{Sha13}
{Shalima}, P., {Jithesh}, V., {Jeena}, K., et~al. 2013, \mnras, 434, 639

\bibitem[{{Shih} et~al.(2011){Shih}, {Charles}, \& {Cornelisse}}]{Shi11}
{Shih}, I.~C., {Charles}, P.~A., \& {Cornelisse}, R. 2011, \mnras, 412, 120

\bibitem[{{Sivakoff} et~al.(2007){Sivakoff}, {Jord{\'a}n}, {Sarazin}
  et~al.}]{Siv07}
{Sivakoff}, G.~R., {Jord{\'a}n}, A., {Sarazin}, C.~L., et~al. 2007, \apj, 660,
  1246

\bibitem[{{Stetson}(1990)}]{Ste90}
{Stetson}, P.~B. 1990, \pasp, 102, 932

\bibitem[{{Tao} et~al.(2011){Tao}, {Feng}, {Gris{\'e}}, \& {Kaaret}}]{Tao11}
{Tao}, L., {Feng}, H., {Gris{\'e}}, F., \& {Kaaret}, P. 2011, \apj, 737, 81

\end{thebibliography}

\end{document}